\documentclass{agujournal2019}
\usepackage{url} 
\usepackage{lineno}
\usepackage{graphicx}
\usepackage{appendix}
\usepackage{verbatim}
\usepackage{tikz}
\usepackage{lineno}
\usepackage{placeins}
\usepackage[utf8]{inputenc}
\usepackage{natbib}
\usepackage{amsmath}
\usepackage[inline]{trackchanges} 
\usepackage{soul}
\usepackage{appendix}

\journalname{JGR: Machine Learning and Computation}

\begin{document}

%
%


\title{Segmentation and Tracking of Eruptive Solar Phenomena with Convolutional Neural Networks}

%
%




\authors{Oleg Stepanyuk, Kamen Kozarev}
\affiliation{}{Institute of Astronomy and National Astronomical Observatory, Bulgarian Academy of Sciences, Sofia 1784, Bulgaria}




\correspondingauthor{Kamen Kozarev}{kkozarev@astro.bas.bg}




\begin{keypoints}

\item{We present our general approach to efficient data-driven solar eruption image segmentation of eruptive solar phenomena.}

\item{We performed a hyperparameter search in order to find the optimal network, and trained U-Net-based models for segmentation of SDO/AIA data.}

\item {We demonstrate the performance of our models on coronal bright front imagery observed in extreme UV data from the SDO/AIA telescope suite.}





\end{keypoints}

%
%

%
%


\begin{abstract}
Solar eruptive events are complex phenomena, which most often include coronal mass ejections (CME), CME-driven compressive and shock waves, flares, and filament eruptions. CMEs are large eruptions of magnetized plasma from the Sun’s outer atmosphere or corona, that propagate outward into the interplanetary space. Over the last several decades a large amount of remote solar eruption observational data has become available from ground-based and space-borne instruments. 
This has recently required the development of software approaches for automated characterisation of eruptive features. Most solar feature detection and tracking algorithms currently in use have restricted applicability and complicated processing chains, while complexity in engineering machine learning (ML) training sets limit the use of data-driven approaches for tracking or solar eruptive related phenomena. Recently, we introduced Wavetrack - a general algorithmic method for smart characterization and tracking of solar eruptive features. The method, based on a-trous wavelet decomposition, intensity rankings and a set of filtering techniques, allows to simplify and automate image processing and feature tracking.  Previously, we applied the method successfully to several types of remote solar observations. Here we present the natural evolution of this approach. We discuss various aspects of applying Machine Learning (ML) techniques towards segmentation of high-dynamic range heliophysics observations. We trained Convolutional Neural Network (CNN) image segmentation models using feature masks obtained from the Wavetrack code. We present results from pre-trained models for segmentation of solar eruptive features and demonstrate their performance on a set of CME events based on SDO/AIA instrument data. 
\end{abstract}

\section*{Plain Language Summary}
Our study focuses on enhancing detection and analysis of imaging observations of solar eruptive events, which is important for predicting space weather impacts on satellites and communication systems. Segmentation and tracking of solar eruptive events based on ground based and remote instrument data is essential in pursuit of better answers about the physical nature of these phenomena. A data-driven approach appears as an obvious solution; nevertheless, engineering of training sets is a challenging task as obtaining labeled ground truth masks requires significant amount of manual work. Existing algorithmic tracking software does not allow to automate the process significantly, as each event or set of the images usually requires separate re-configuration of the software. We used our previously developed wavelet-based algorithmic method, which to a certain extent simplifies the segmentation of solar images, together with synthetic data for engineering of training sets. We describe several image segmentation models and demonstrate their performance, and generally discuss various aspects of high-dynamic range images segmentation in heliophysics domain.

\section{Introduction}
\label{introduction}

Coronal Mass Ejections (CMEs) are large expulsions of plasma and magnetic field from the solar corona into the heliosphere. CMEs travel outward from the Sun at speeds ranging from slower than 250~km/s to more than 3000~km/s \citep{Yashiro:2004}. Due to their ability to generate large Solar Energetic Particle (SEP) events and severe geomagnetic storms, CMEs are regarded as a significant natural hazard that can cause disruptions in the interplanetary environment, with potential impacts on spacecraft, satellites, and communication systems.

Since their discovery, CMEs have been associated with a host of solar activity processes \citep{Hansen:1971,Tousey:1973}. They are known to be formed by explosive reconfiguration of solar magnetic fields through the process of magnetic reconnection, but broad theoretical discussion aiming to shed light on the exact formation mechanism still continues \citep{Shibata2011, Chen2011}. To accurately identify and measure the size of the erupting bubble and the subsequent flux rope, high-cadence, high-sensitivity off-limb extreme ultraviolet (EUV) and white light observations have proven very useful \citep{Carley:2020, Jiang:2021}. Due to their increasingly large volume, these observations must be accurately processed which sets demands for development automated pipelines and more advanced interpretation techniques.

The launch of the Solar Dynamics Observatory \citep[SDO]{Pesnell:2012}, with its high-resolution, high-cadence EUV telescope Atmospheric Imaging Assembly \citep[AIA]{Lemen:2012}, has enabled even more detailed studies of CMEs \citep{Kozarev:2015} in the lower corona, as well as extensions to modeling conditions for coronal acceleration of solar energetic particles in compressive fronts \citep{Kozarev:2016, Kozarev:2017, Kozarev:2019}. However, the complexity of CMEs and related phenomena still presents challenges for precise instrument design and data interpretation \citep{Wuelser:2004,Barnes:2019}. One of these phenomena are EUV waves, also known as coronal bright fronts (CBF) - relatively dim, large-scale traveling disturbances, observed both on disk and off limb. These are thought to correspond to compressive and shock waves in the low and middle solar corona, and have been previously observed in EUV wavelengths \citep{Kozarev:2010, Veronig:2010}.


Automatic algorithmic or data-driven differentiation between eruptive events in the solar corona can be challenging due to the complex interplay of phenomena in the Sun's atmosphere. When visually examining observations, other contextual information, such as the presence of a solar flare or coronal mass ejection, can provide clues about the likelihood of observing one of these phenomena. Additionally, working with observations at multiple wavelengths and using time-lapse imagery to observe the propagation and evolution of these structures can further assist in analysis \citep{Stepanyuk:2024}.

In recent years, a few groups independently proposed a number of algorithmic, non-data driven approaches for EUV wave detection of varying complexity \citep{Podladchikova:2005, Verbeeck:2014, Long:2014, Ireland:2019, Stepanyuk:2022}. Of these perhaps the most advanced are the CorPITA \citep{Long:2014} and AWARE \citep{Ireland:2019}. The CorPITA algorithm employs percentage base difference (BD) images to fit multi-Gaussian shapes to flare-source-centered, sector-averaged portions on the wave along the solar disk. The AWARE algorithm uses more advanced pre-processing in persistence running difference images, to characterize the wave shapes along similar flare-centered sectors, and a random sample consensus (RANSAC) algorithm to select the features. However, using running difference images introduces spurious features and is not recommended for discovering the true (projected) wave shape \citep{Long:2011}. In addition, the persistence imaging approach tends to amplify the noise in the output, requiring additional post-processing (median filtering and closing operations). While both these algorithms are well automated based on flare onset signal, they have so far been applied only on the solar disk. What is more, their procedures are focused specifically on EUV waves.

In \citet{Stepanyuk:2022} we introduced Wavetrack, a wavelet-based algorithmic framework for solar feature detection and tracking. Recently we presented an updated version which was applied to four different instruments, to study a single CME extending from the solar surface to more than 20 solar radii.  The computation scheme is based on a multi-scale data representation concept \citep{Starck:2002}, combined with an à trous wavelet transform \citep{Akansu:1991, Holschneider:1989, Stenborg:2008} and a set of image filtering techniques. The proposed method has allowed, to a certain extent, to automate feature recognition for multiple events. For more details on the method, please see \citet{Stepanyuk:2022, Stepanyuk:2024}.

Most of the algorithms currently used in solar feature detection and tracking can be helpful for developing sufficient training sets for applying a modern artificial intelligence approach to the problem. Nevertheless, the complexity of their processing chains makes them difficult to implement. In addition, they are usable for very specific tasks, and are often applicable only to data from a single instrument. Specifically, the task of EUV wave recognition and tracking is complicated by their much weaker intensity compared with most other solar features, next to which they propagate and project.

Generally, better image segmentation techniques for (CBF) tracking enable precise mapping of their morphology and boundaries, revealing fine-scale structures that helps to test theories of coronal wave propagation. Enhanced tracking and velocity measurements provide accurate kinematics, including deceleration profiles and associations with flares or eruptions, can be fed into numerical MHD models as boundary conditions or validation data to simulate energy transfer in the solar corona. Improvements constrain competing theories by offering statistical limitations on wave damping, reflection, and dissipation mechanisms, reducing uncertainties in solar atmospheric dynamics.

In recent years Machine Learning (ML) and Deep Learning (DL) methods have become more frequently applied to specific solar physics problems, as they are well suited to extracting useful information from time series and large multi-wavelength imaging data. For instance, \citet{Szenicer:2019} used a convolutional neural network to produce EUV irradiance maps from AIA images. \citet{Li:2013} proposed a multi-layer model to predict solar flares based on sequential sunspot data; and \citep{Kim:2019} applied generative adversarial networks (GAN) to generate the magnetic flux distribution of the Sun from SDO/AIA images. \citet{Nishizuka:2018} tested three different machine learning algorithms – support vector machine (SVM), k-nearest neighbors (k-NN), and multi-layer perceptron (MLP) – for predicting solar flare occurrences within the next 24 hours.


Looking at the progress in neighboring scientific fields, convolutional neural networks (CNN) intuitively seem to be the first choice technology for tracking of CME-related phenomena (\citet{Stuardi:2024, Panes:2021, Hausen:2020, Zhao:2025})  still, application of CNNs for image segmentation in heliophysics has been limited. Sufficient training sets for segmentation tasks in imaging observations could consist of thousands of images, depending on the choice of CNN model and training strategies. Engineering of a training set used to be a challenging task, as existing algorithmic (non-data driven) tools mostly lack ability to process multiple events within single configuration/setup and are thus ill suited for use in pipelines. Processing of great amounts of data through manual detection and labeling of features is extremely time consuming, while preparing pixel-segmented image data for ground truth is even more difficult. Application of our wavelet-based image segmentation method Wavetrack allows to simplify this process and put emphasis on training data quality and its sufficiency at a new level.

In this work, we describe and demonstrate CNN-based methods for hybrid algorithmic - data-driven characterization and tracking of solar observational features, focused on coronal bright fronts. After feature masks are obtained with Wavetrack, engineered training sets are passed through the CNN architecture. In a series of supervised training rounds within a general hyper-parameter \citep{Feurer:2019} search routine, optimal CNN model parameters (width, depth, filters) are determined in order to provide the best performance for the selected instrument and/or data type (i.e base difference, running difference or raw data).

We find that models developed within our approach perform at least as well as the control results obtained by Wavetrack. Once trained, they can be efficiently applied to a larger number of EUV wave events. This is a step forward towards automatic data-driven segmentation in the heliophysics domain, as trained CNNs are largely configuration‑free models that scale across many events and integrate cleanly into automated pipelines for a given instrument. Although training is computationally costly, inference is lightweight: pre‑trained models generate masks efficiently and can run on modest or embedded hardware, which is advantageous for onboard processing in space missions. The models are regularly improved and updated, and new channels and instruments will be added to the package in a short-middle term perspective.

The paper is structured as follows: The method section \ref{method} starts with discussion on general concepts of data-driven segmentation of eruptive solar phenomena. It is followed by description of our approach towards processing of the input data, engineering of a training set and data augmentation, given in section \ref{trainset_engineering}. We discuss models training and their performance in the section \ref{results}. 

%






\section{Method}
\label{method}

\subsection{Data driven image segmentation: The general concept}
\label{image_sementation_general}

Image segmentation is a computer vision technique used to divide an image into multiple segments or regions based on certain criteria, such as texture, intensity, or other visual features. The goal of image segmentation is to simplify or change the representation of an image into more meaningful and easier-to-analyze parts. The output of an image segmentation algorithm is a set of labeled regions or objects that represent the different parts of the original image. Technically, this means assigning a corresponding category of each pixel, i.e if there exist two objects of the same category on the input image, the segmentation map does not inherently distinguish these as separate objects. 
For image segmentation of CBFs, we have searched for a DL model, which would perform  well for this type of data (AIA 193\AA~images, which have low contrast, high noise levels), and at the same time require a relatively small size of the training set. 

\begin{figure}[htp]
    \centering
    \includegraphics[width=12cm]{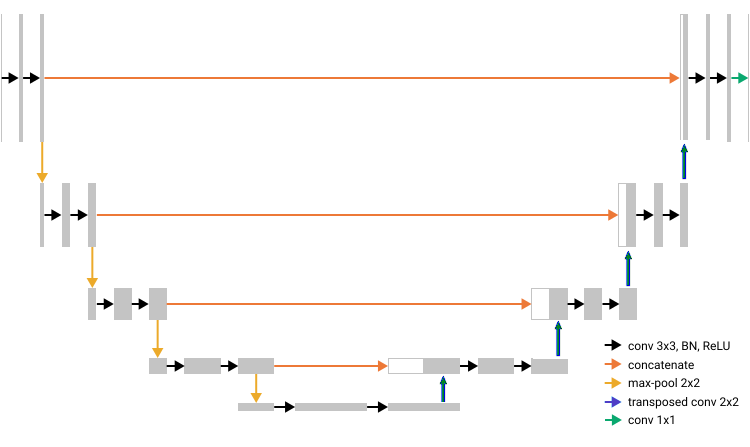}
    \caption{U-NET architecture in general.  Each gray box corresponds to a multi-channel feature map. (The number of channels is not specified, as it would depend on the network setup in each specific case) }
   \label{fig:fig_unet_fig_1}
\end{figure}

We decided to base our approach on the U-Net architecture\citep{Ronneberger:2015}, which captures its architecture of a contracting path followed by an expansive path, connected by a central bottleneck (Fig. \ref{fig:fig_unet_fig_1}).  Its efficient and effective design has led to its adoption in a variety of domains, from satellite imagery analysis to industrial defect detection. In the contracting or encoder path, the network starts with an input image and applies a series of convolutional layers, each followed by a rectified linear unit (ReLU) activation. These convolutional layers are then followed by max-pooling layers that down-sample the image. This process effectively captures the contextual information of the image while simultaneously reducing its spatial dimensions, albeit increasing its depth, meaning the number of feature maps. The center of the U, known as the bottleneck, consists of two $3\times3$ convolutions each followed by a ReLU layer, without any subsequent max-pooling. This layer sits between the contracting and expansive paths and captures the most compressed representation of the input image. The expansive or decoder path of the network up-samples the feature maps. 

In each step of this path, the network upsamples the feature map, performs a 2$\times$2 up-convolution to increase its spatial dimensions, and then concatenates it with a correspondingly cropped feature map from the contracting path. After this concatenation, two 3$\times$3 convolutions followed by ReLU activations are applied. These connections from the contracting path, often referred to as `skip connections', help the network retain spatial information that might otherwise be lost during the downsampling process in the encoder.
The architecture culminates in a final 1$\times$1 convolution layer that maps each feature vector to the desired number of classes, producing the segmented output. One of the standout features of U-Net is its skip connections that bridge the encoder and decoder paths. These connections are crucial in preserving the finer details, vital for precise segmentation tasks. Another advantage of the U-Net architecture is its relatively fewer parameters compared to other similar architectures. This not only makes it faster to train but also makes it particularly suitable for scenarios where the amount of available training data is limited. Our general approach to CNN design, is not tightly bonded to a specific CNN architecture like U-NET which was chosen for the current research and allows to explore other architectures in the future (see hyper-parameter search). 

\subsection{Hyper-parameter search. Our approach to CNN design}
\label{hyper-parameter}

Designing a U-Net (and any CNN) architecture for specific cases involving high dynamic range images with features located within a narrow intensity range and relatively large feature sizes requires careful consideration of both the network's depth and width. Since image features of interest are relatively large, a very deep network might not be necessary. If the task involves large structures, a shallower U-Net may suffice, as deep networks are typically more useful for capturing fine, intricate details, which are less important in our scenario. Nevertheless, High dynamic range (HDR) images, such as those from AIA, are complex and require a network with greater capacity to adequately capture the necessary feature representation. This suggests the use of a wider network, meaning more filters per layer. At the same time, larger filters (5$\times$5 or even 7$\times$7) can capture more spatial information in each convolution operation. With larger filters, each convolutional layer can cover a larger receptive field, potentially reducing the need for very deep networks to capture the necessary spatial context. Larger filters might simplify the network architecture, as they process a larger portion of the image in each operation.

Given the high dynamic range, batch normalization helps stabilize and speed up training by normalizing layer inputs. It standardizes activations (zero mean, unit variance) and then scales and shifts them using learnable parameters (gamma and beta). This reduces internal covariate shift, enables higher learning rates, and acts as a regularizer \citep{hinton:2012,srivastava:2014}. Dropout is a regularization technique that prevents overfitting by randomly disabling a fraction of neurons during training. This encourages the network to learn redundant, robust features that generalize better. We have chosen empirically dropout rates ranging from 0.2 to 0.4. Dropout effectively simulates training multiple networks with minimal extra cost \citep{ioffe:2015,ioffe:2017,santurkar:2018}.

We test our general suggestions about the model architecture within hyper-parameter search paradigm. Hyper-parameter search is a crucial step in machine learning and deep learning to fine-tune the configuration of models for optimal performance. Hyper-parameters are the configuration settings used to structure the machine learning model, which can significantly influence the behavior and performance of the model. Unlike model parameters that the model learns from the training data (e.g., weights), hyper-parameters are set prior to the training process and remain constant during training.There are several methods to search for the best set of hyper-parameters, with grid search and random search being among the most popular. Grid search involves defining a grid of hyper-parameters and systematically working through multiple combinations of these hyper-parameters. We specify a list of values for each hyper-parameter (model depth, start filters count, filter size, intensity augmentation rate) and evaluate the model performance for each combination of these hyper-parameter values, training the model each time and assessing its performance using a predefined metric - correspondence with a validation set data.

We tested two metrics - the Intersection over Union (IoU), also known as the Jaccard index, and the Dice coefficient (also known as the Sørensen–Dice coefficient) to quantify the similarity between two images. IoU is defined as the size of the intersection divided by the size of the union of two sets. In the context of comparing two images or segmentation masks, it is calculated as:
\\
\text{IoU}(A, B) = $\frac{|A \cap B|}{|A \cup B|}$ 
\\
The Dice coefficient is similar to IoU but it gives more weight to the intersection part. It is defined as twice the size of the intersection divided by the sum of the sizes of the two sets:
\\
\text{Dice}(A, B) = $\frac{2 \times |A \cap B|}{|A| + |B|}$
\\
The Dice coefficient is more sensitive to the size of the overlap than IoU. It tends to give higher values compared to IoU for the same amount of overlap due to its formulation. It appears to be more sensitive to small structures. This can be advantageous for filaments due to their thinner shape, but for comparing predicted and validation set CBFs in this paper we used IoU coefficient.

\section{Input data. Training sets engineering}
\label{trainset_engineering}

The engineering of a training set in CNN  model training refers to the process of selecting and preparing a set of training data to train the model.  The quality of the training data are crucial factors in determining the accuracy and generalization performance of the CNN model. The training data should be diverse, representative, and balanced to ensure that the model can learn the underlying patterns and features of the data and avoid overfitting. Training sets were generated by applying Wavetrack to a serious of eruptive events observed by the AIA telescope. Part of the events were previously studied by \citet{Kozarev:2015, Kozarev:2017, Stepanyuk:2022, Stepanyuk:2024, Rigney:2024}.

The computational scheme goes as follows: The data is processed to level 1.5 using the standard processing pipeline available in aiapy \citep{Barnes:2020} and SunPy \citep{Sunpy:2020} libraries. The BD images are obtained by subtracting a base image from each timestep of the image data. Using BD allows to enhance the change in intensity, caused by the eruptive front, omit static details, and reduce noise. Base images are created as an average of several (3-5) consecutive timesteps 2-5 minutes prior to the beginning of the event in each instrument. Absolute values of the threshold interval (typically found by us to be in the range [-50,150] for AIA BD data) are selected for the purpose of selecting of  a relevant part of the image dynamic range. At the next step, BD images are decomposed with the \`a trous wavelet technique into a series of scales, with third and fourth wavelet scales chosen for image recomposition. This choice is empirical and is based on our experience in processing AIA observations of EUV waves.

The  \`a trous wavelet technique involves a recursive algorithm that performs a series of convolutions and sub-sampling operations on the image. This decomposition creates a multi-resolution representation of the image, with different levels of detail at different scales. The technique has several advantages over other wavelet transform methods. It has a computationally efficient algorithm, requires fewer wavelet coefficients, and can produce better results for certain types of images, such as images with details with distinct edges. Calculation of Sobel-Feldman gradient field of the reconstructed image is used to further sharpen the edges. To each of the wavelet coefficients a relative thresholding is applied once more depending on the statistical distribution of the pixel intensities for each of the decomposition levels. Stand-alone masks of every object are obtained via a segmentation routine. As the final step, we have added a noise filter to Wavetrack. It is a single pass window routine, which applies a filter to an image by examining each pixel and its surrounding neighbors within a certain window size (it may vary from event to event, but typically 20$\times$20 pixels for removal of noise and spurious small-scale features). If the difference between the pixel and its neighbors is below a certain threshold value, the pixel is modified to be the average of its neighbors. If the difference is above the threshold, the pixel is left unchanged. This filter can effectively remove smaller details from an image while preserving the larger structures and features.

\begin{figure}[htp]
    \centering
    \includegraphics[width=16cm]{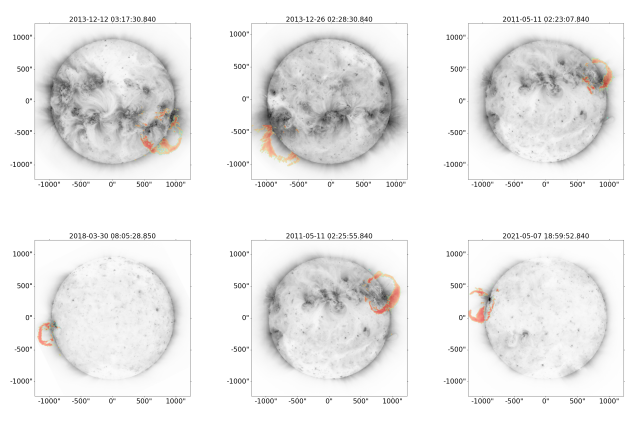}
    \caption{
     Five separate example CBF events from the initial dataset used for engineering of training set. May 11, 2011; June 07, 2011; December 12, 2013; December 26, 2013, May 07, 2021. Wavetrack-obtained masks were projected over inverted-colormap grayscale AIA 193\AA~images. Events and timesteps were selected to represent fronts of different shapes in various heliospheric latitude-longitude zones for the purposes of creating diverse training set with further engineering of a synthetic training data through data augmentation routine.}
    
    \label{fig:wavetrack_multievent}
\end{figure}

Figure \ref{fig:wavetrack_multievent} illustrates the objects successfully extracted by Wavetrack for a few events selected from the full set of events we chosen for the training set engineering. Initial set of events and time-frames was chosen to represent  fronts of various shapes in various heliospheric latitude-longitude zones. The dataset is  partitioned into training, validation, and testing sets. The training set is used to train the model, while the validation set is used to monitor the model's performance during training and adjust the model's parameters accordingly. The testing set is used to evaluate the final performance of the trained model on unseen data. 

Potential limitations of using Wavetrack output for training data may result due to imperfections of algorithmic (wavelet-obtained) masks. We recognize that training an image segmentation (CNN) model with imprecise or imperfect masks can significantly impact model performance, resulting in reduced segmentation accuracy due  to a standard set of issues such as overfitting to noise, increased false positives or negatives, poor generalization, or training instability. To which extent this problem will show up in our case depends on how well Wavetrack was configured in each case. We have aimed to minimize such problems by pre-selecting image-mask pairs from the series of events with the most clean basic data possible.

\subsection{Data Augmentation. Synthetic Data}
\label{Data_Augumentation}
Data augmentation is a technique used in machine learning and computer vision to increase the size of a training dataset by applying various transformations to the original data. One common method of data augmentation is rotation, where the images are rotated by a certain degree to create additional training examples. In the case of image data, rotation augmentation involves rotating the image by a certain angle and then padding the areas that become empty as a result of rotation. For example, if an image is rotated by 45 degrees, the corners of the original image will become empty, and padding will be needed to fill in those areas.

\begin{figure}[htp]
    \centering
    \includegraphics[width=16cm]{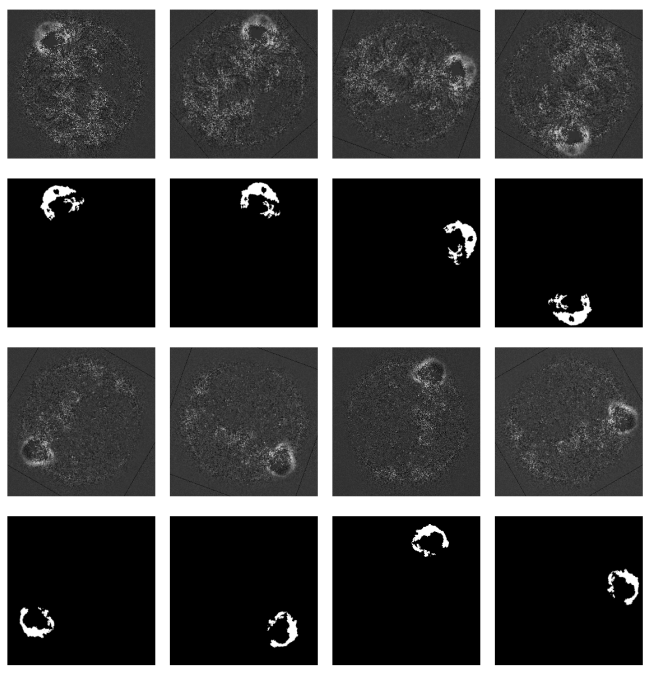}
    \caption{Rotation argumentation example for the two events from the training set. Rotated BD images are shown on the first and third row, with corresponding feature masks (ground truth) in the second and forth rows, respectively. Columns correspond to random angles chosen by the algorithm during one of the training steps on synthetic data.}
    \label{fig:fig_augumentation}
\end{figure}

The purpose of data augmentation with rotation is to help the model learn to recognize objects from different viewpoints and orientations, which can improve its ability to generalize and make accurate predictions on new, unseen data. It also helps to reduce over-fitting, as the model is exposed to more diverse examples of the same object. We used 5 degree rotation step of every BD-mask pair from the initial dataset for training of our model (Fig. \ref{fig:fig_augumentation}).

Intensity Augmentation specifically refers to altering the pixel intensities of the input images. In case of BD images from the training set used in this work, only areas limited by the actual feature shape were augmented.  That was done by projecting the feature mask (Fig  1. left image) to the BD image with future variation of the pixel intensity in [-0.5,0.5] interval with 0.1 sigma standard deviation increment, assuming pixel intensities are following Gauss distribution. Depending on the model, training set size together with synthetic augmented data can be up to a few tens of thousands of dynamically generated images.

\subsection{Sequential training and data augmentation}

Recently, sequential training strategies in deep learning have garnered significant interest, with curriculum learning (CL), gradual exposure, and hybrid augmentation methods offering systematic improvements over traditional training paradigms. CL, particularly for image segmentation tasks using architectures like U-Net, involves structuring training data presentation from simpler to more complex examples, often integrated with data augmentation to enhance model generalization on limited datasets. This approach legitimizes gradual exposure to variations, such as incremental transformations, by allowing the network to first master core patterns before adapting to distortions, thereby reducing overfitting and improving robustness. A recent study demonstrated superior performance over standard random augmentation in terms of accuracy and stability \citep{choi:2024}. CL formalizes the intuitive practice of training models from simple to progressively complex data. A critical element of CL involves defining appropriate difficulty metrics, such as semantic similarity \citep{Kang:2020}, or structural hierarchies \citep{yang:2025}. Dynamic scheduling represents another advancement, where adaptive pacing mechanisms adjust the learning trajectory based on real-time model competence. Similarly, progressive curriculum learning with a Scale-Enhanced U-Net stages feature extraction from coarse to fine scales, yielding better segmentation continuity and efficiency compared to classical U-Net. These techniques are particularly beneficial for small datasets, offering noticeable gains in metrics like IoU by mimicking human learning paradigms and optimizing convergence.

Sequential data augmentation methods further embody the CL philosophy by incrementally introducing data complexity. It extends curriculum principles through phased retraining, where models are iteratively exposed to augmented data in cycles, fostering gradual adaptation without overwhelming the network early on. This method supports gradual exposure by sequencing augmentations, such as rotations or deformations. Gradual augmentation protocols, such as rotation increments of a certain angle (often 5°), function as implicit regularizers, mitigating overfitting risks, particularly in small datasets ($<$100 images). \citet{yang:2025}] illustrated how phased augmentation stabilizes gradient variance more effectively than one-shot augmentation. Nevertheless, incremental approaches entail linear increases in computational cost proportional to augmentation phases (e.g., 72 increments for full 360° rotation), potentially yielding diminishing returns. Strategies to alleviate computational overhead, such as entropy-based pruning of redundant transformations, have been explored by \citet{CHEN:2024}.

Hybrid methods integrate phased augmentation with architectural adaptations, optimizing performance. Scale-Enhanced U-Net, for instance, incorporates curriculum staging with multi-scale input handling, effectively addressing intra-class imbalances and notably enhancing Dice scores for small-structure segmentation \citep{yang:2025}. Conversely, `orthodox' augmentation techniques, such as single-phase elastic transformations originally introduced with U-Net, prioritize computational efficiency but may suffer from mode collapse in highly variable datasets. Classic augmentation strategies, which apply random or fixed transformations without curricular ordering, offer simplicity and efficiency but may under-perform in scenarios with extreme data scarcity or high variability and encounter difficulties managing hierarchical features compared to curriculum-based approaches \citep{Liu:2021}.

In summary, curriculum learning and sequential augmentation methodologies offer demonstrable advantages, particularly within low-data scenarios and tasks featuring complex hierarchical features. A trade-off between efficiency and performance emerges distinctly within these methodologies. In our work we use both approaches - `classic' training and augmentation, and curriculum learning and sequential augmentation.

\section{Model training and performance on image segmentation tasks}
\label{results}

We start with a baseline architecture (64 filters in the first layer, 3$\times$3 filter size, 3 layers depth, 0.1 standard deviation pixel intensity augmentation) and adjust depth, width and filters size/count based on models performance (see hyper-parameter search). The synthetic augmented training set is generated dynamically and used alongside the original training set.

\begin{figure}
    \centering
    \includegraphics[width=16cm]{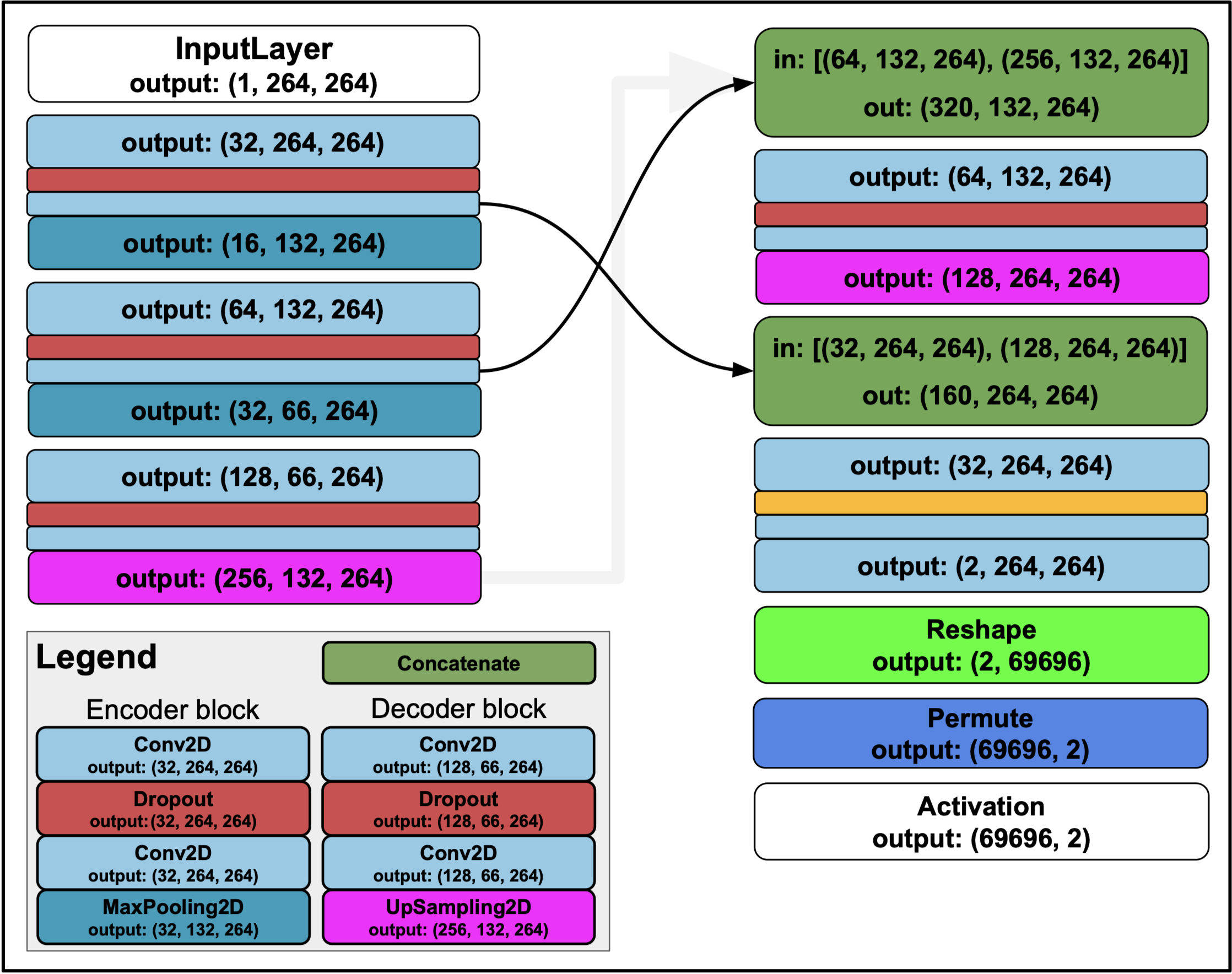}
        \caption{Compact CNN model structure. It is based on a hyper-parameter search based on segmentation performance evaluation on BD images, [-50,150] absolute intensity threshold interval applied. Due to its vertical size the plot was split vertically into 2 columns for the purpose of more compact visual representation. Thin black arrow lines denote skip connections. The legend below the architecture describes the structure and dimensionality principle of the encoder and decoder blocks used.}
    \label{fig:compact_model}
\end{figure}

\begin{figure}
    \centering
    \includegraphics[width=16cm]{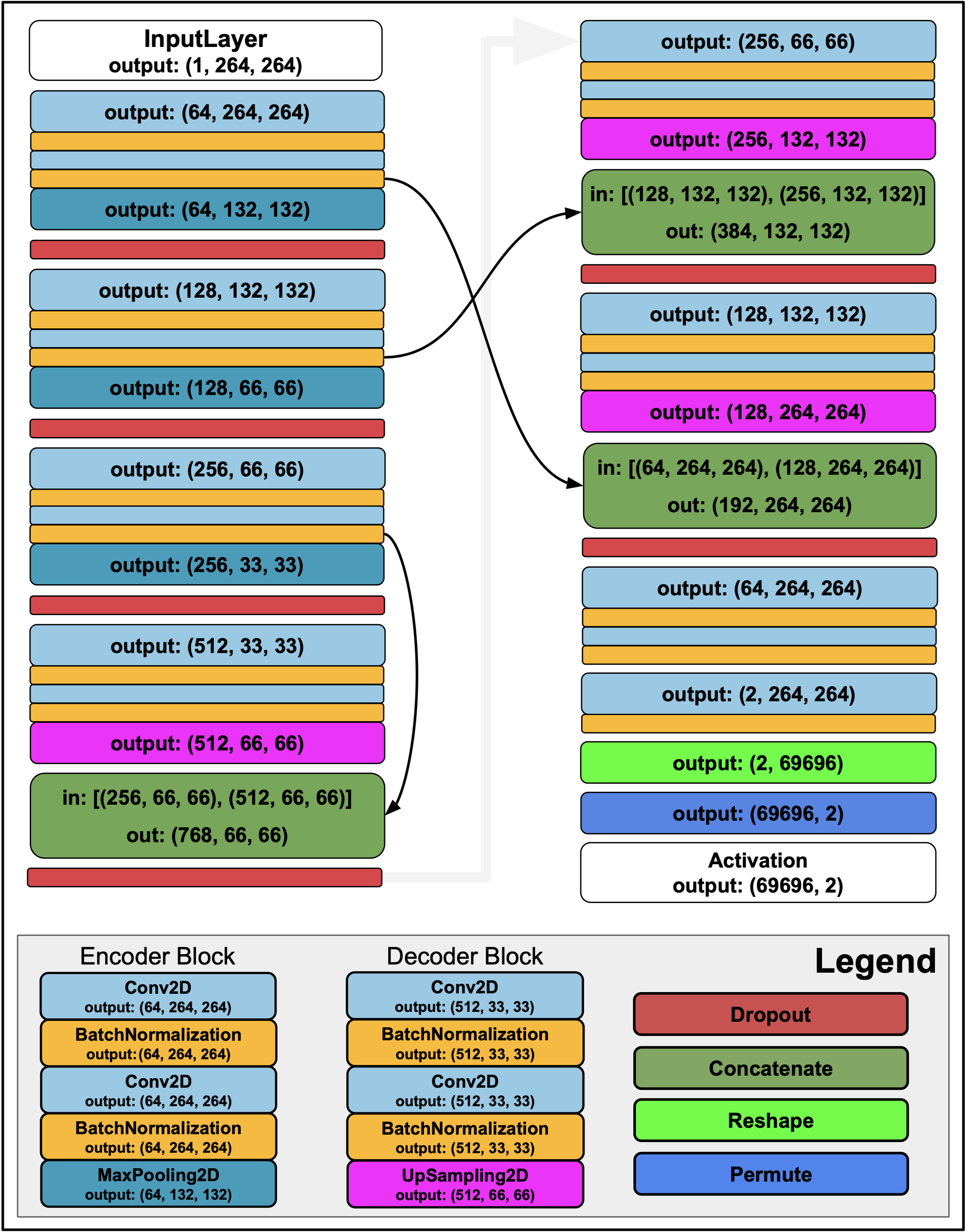}
        \caption{Large CNN model structure. It is based on a hyper-parameter search based on segmentation performance evaluation on BD images with no threshold applied. This more complex and deep network structure was generated as segmentation of a higher dynamic range images was attempted. Due to its vertical size the plot was split vertically into 2 fragments for the purpose of more compact visual representation. Thin black arrow lines denote skip connections. The legend below the architecture describes the structure and dimensionality principle of the encoder and decoder blocks used.}
    \label{fig:big_model}
\end{figure}


It is not always advantageous to train a model for the maximum number of epochs. While more epochs can mean more learning from the training data, this does not necessarily translate to better performance on unseen data. Training for too many epochs can lead to overfitting, where the model learns the noise and details in the training data to an extent that negatively impacts its performance on new data. 

To optimize the training duration without compromising the model's ability to generalize, implementing an early stopping technique is beneficial. Early stopping involves monitoring a chosen performance metric on a validation set at each epoch and stopping the training when this metric stops improving or begins to degrade. 

In our approach, non curriculum models are iteratively trained until average 0.7 correspondence of predicted masks with masks from independent validation set generated algorithmically by Wavetrack. The IoU coefficient was used as a comparison metric for the masks representing CBFs. Every few hundred of epochs the training routine is paused to perform interim model performance evaluation based on comparison with the validation set. This concept was technically implemented through TensorFlow call-back functions. Mentioned above CNN parameters such as depth, width, and the size and quantity of filters, are intricately linked to the dynamic range of the dataset on which it is trained. Efficiency of the training routine depends on balancing of interplay between these parameters.

In Figure \ref{fig:big_model} we show the structure of a large model generated from hyper-parameter search on non-thresholded data. Higher dynamic range in the data, indicating a wide spectrum of intensity values, demands a network with sufficient depth and width to effectively capture the detailed features at multiple scales. To counterbalance the increased risk of overfitting associated with deeper and wider networks,  expanding the training dataset size becomes essential. Larger datasets provide a more varied set of examples, which helps in generalizing the model better by reducing the model’s capacity to memorize noise and specific training samples. We address this issue by more extensive use of dynamically generated synthetic data reducing augmentation angle to 1 degree in case of HDR data.

\section{Model performance on validation sets}

\begin{figure}[htp]
    \centering
    \includegraphics[width=16cm]{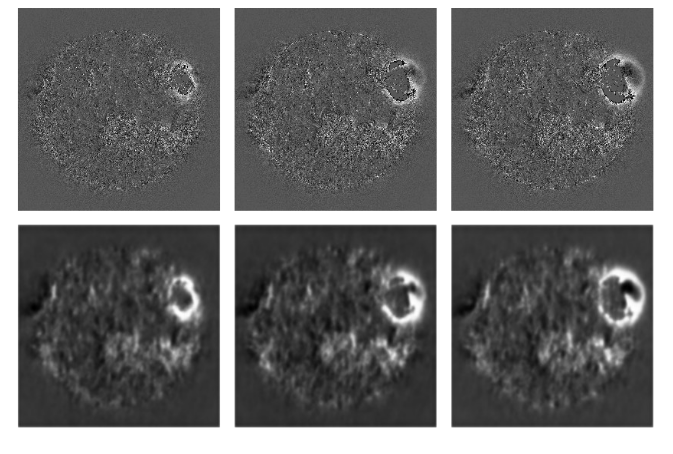}
    \caption{
     Feature segmentation results, probabilistic masks (bottom row), for a random event from the validation set. Upper row shows corresponding BD images (after thresholding).}
    
    \label{fig:results_BD_pb}
\end{figure}

First type of models we present (Fig. \ref{fig:compact_model}) are designed to predict the likelihood of each pixel belonging to a certain class or feature in the image (Fig. \ref{fig:results_BD_pb} and \citet{Stepanyuk_dataset:2024}). This is fundamentally a probabilistic task, where the model assesses the probability of each pixel's membership in a given class. Instead of just assigning a pixel as `part of a feature' (1) or `not part of a feature' (0), the model provides a continuous probability score between 0 and 1. This approach offers more nuanced information about the model's confidence in its predictions.

The final layer employs a  softmax (for multi-class segmentation) or sigmoid (for binary segmentation) activation function. These functions are crucial for converting the raw output of the neural network into probabilities. While a sigmoid function maps any input into a value between 0 and 1, a softmax function does the same but also ensures that the sum of probabilities for all classes at each pixel equals 1, making it suitable for multi-class problems. 

By outputting a range of values, the CNN provides a more detailed and flexible interpretation of the image. A probabilistic output allows for a more nuanced interpretation of ambiguous region. The continuous output can be converted into binary masks through thresholding. By choosing a probability threshold, pixels with probabilities above this threshold are classified as part of a feature, while those below are not. This threshold can be adjusted based on specific requirements or desired sensitivity, providing flexibility in how the results are interpreted and used. The range of values also gives insights into the model's confidence. Areas where the model is unsure will have probabilities around 0.5, indicating ambiguity, probabilities below 0.5 indicate that it is unlikely that pixels belong to the feature of interest.

\begin{figure}[htp]
    \centering
    \includegraphics[width=16cm]{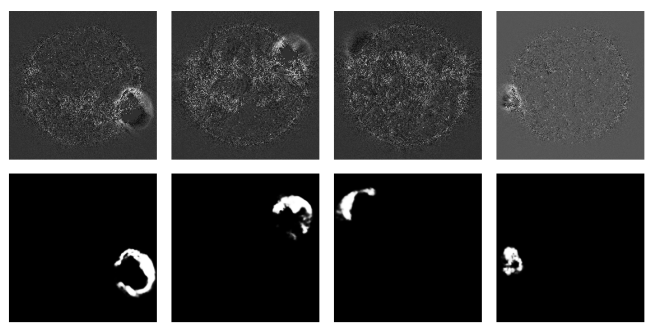}
    \caption {Feature segmentation results, binary masks (bottom row). Upper row shows corresponding BD images. Events and timesteps were selected from the validation set to represent fronts of different shapes in various heliospheric latitude-longitude zones. }   

    \label{fig:results_BD}
\end{figure}

On Fig. \ref{fig:results_BD} we present segmentation results from the model designed to produce binary masks. The upper row shows random BD images from random events from the validation set. On the bottom row binary masks captured by the model are shown. Binary masks reveal CBFs reasonably well on the limb part of the image, giving somewhat less flexible results than a probabilistic model.  \citep{Stepanyuk_dataset:2024}

On-disk and off-limb parts of images have significantly different statistical distributions of pixel intensities, thus may often require different sets of parameters for filtering and decomposition techniques for algorithmic, non-data driven approaches \citep{Stepanyuk:2022, Stepanyuk:2024}. From the point of image processing on-disk and off-limb parts can be considered as essentially two different types of images.
Same logic can be somewhat extended into data-driven image segmentation realm. In the current study, large part of the of images from training sets had features which were located off-limb. For the models presented by the time of this publication, under-representation of on-disk features in training sets may result in worse performance for on-disk parts of the images and false-labeling brighter areas as features 
which sets demands for future improvements of the models. 
It also makes sense to suggest that more compact models like the one from Fig. \ref{fig:compact_model}, trained on BD images with an intensity threshold, may perform poorly on BD images without somewhat similar threshold intervals applied due to higher dynamic range of such images. We discuss these issues and possible solutions in in Section \ref{case_study}

\section{Case Study: CME of 10 May, 2022}
\label{case_study}

\subsection{The event}

We have applied one of our models to study the CBF in the eruptive event of 10 May, 2022, which stands out as a significant event early in Solar Cycle 25. An X1.5-class solar flare began at 13:50~UT from NOAA active region AR 13006 in the south central region of the solar disk as observed from Earth. It induced pronounced ionospheric disturbances, observable through variations in total electron content (TEC) derived from GNSS data \citep{Lopez-Urias:2023}. These disturbances underscore the profound impact of X-class solar flares on Earth's ionosphere, with implications for space weather prediction and technology-dependent systems.

\citet{Kosovichev:2023} identified rapid continuum emission variations and transient emissions in the flare line core, indicative of intense, impulsive heating in the Sun's lower atmosphere. This analysis revealed a rare and compelling sunquake phenomenon related to the flare, providing a deeper understanding of energy release during flares. \citet{Karlicky:2024} examined narrow-band decimetric radio spikes during the flare, finding that these spikes occurred in quasi-periodic groups with a period of 5.1 seconds, providing insights into the flare's radio emission characteristics.

Further analysis by \citet{Rigney:2024} focused on the dynamics of the flare- and sunquake-associated shock wave and its relation to radio emissions. Using the CORPITA and Wavetrack software, the authors tracked the propagation of a large CBF along a channel in the low solar corona. The motion was linked to the generation of radio bursts observed during the event, highlighting the shock's role in exciting energetic particles. The velocity of the shock wave, estimated between $500$ and $1000~km/s$, provided key data on the kinematics and energy  transport processes during the eruptive event.

\subsection{Data driven image segmentation of the on-disk event}


Both data-driven and traditional algorithmic approaches, particularly convolutional neural networks (CNNs), benefit from having more uniform ranges of pixel values. With fewer large or drifting offsets, they can more reliably learn to detect evolving features rather than grappling with large changes in global intensity. The running difference method helps maintain more consistent intensity changes from frame to frame, because each difference is bounded by similar overall brightness scales. In contrast, base difference images introduce additional challenges due to their cumulative nature.

\begin{figure}[htp]
    \centering
    \includegraphics[width=16cm]{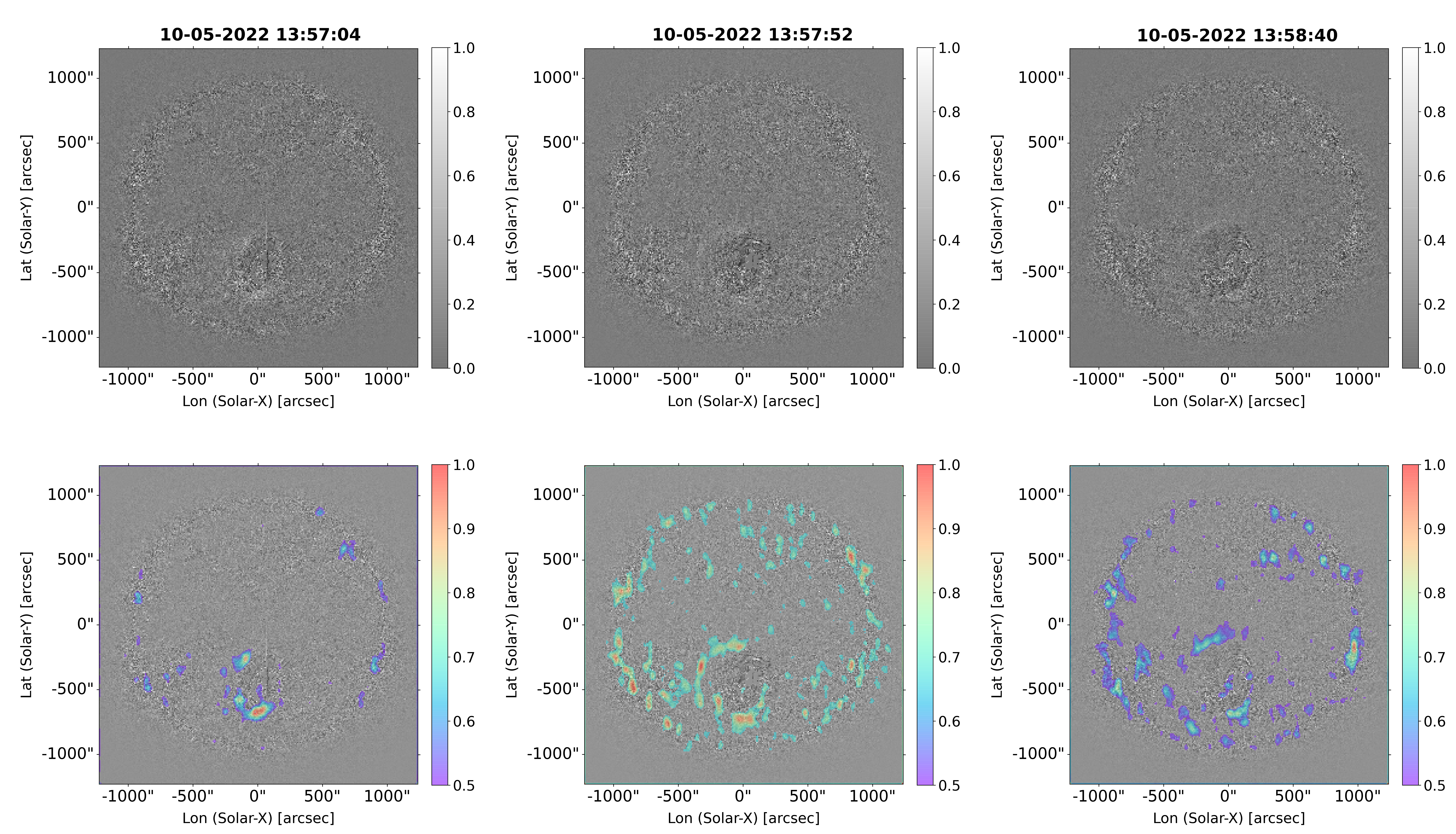}
    \caption {Data-driven image segmentation model output for the May 10, 2022 EUV wave event observed by SDO/AIA (bottom row). The intensity of each pixel corresponds to the probability that this pixel belongs to the wavefront as determined by the model. Input Running Difference images are shown on the upper row.}
    \label{fig:prediction_panel}
\end{figure}

On the Figure \ref{fig:prediction_panel} we demonstrate the performance of one of our models applied May 10, 2022 SDO AIA Running difference data. In the top row we show three input running difference AIA images separated by 48 seconds.

The CBF expands away from the source in the south. The front of the circular wave is slightly discernible. The bottom row shows the hybrid model's probabilistic output. The oval shape of the CBF is clearly visible, split into a northwest and a southeast front. This is best visible on the rightmost image of Fig. \ref{fig:prediction_panel}.

The model used for this case-study was trained in rounds using a sequentially augmented training set consisting of 12960 264$\times$264 pixel images in total. The engineered synthetic training set is based on initial 180 pairs generated by Wavetrack, which were pre-selected from the series of events aiming to have the most diverse, representative and clean basic data. For our models, we usually chose the minimum effective image size (and therefore resolution) that still allows the morphology and dynamics to be resolved with satisfactory precision. That allows to reduce computational resource usage, while also maintaining the potential for spacecraft on-board software applications of such models.


\subsection{Velocity field estimation}
In solar physics, the (FLCT) method is often considered to be the first-choice technique for determining horizontal flows, particularly in active regions. The  method calculates local shifts between two images in the Fourier domain, which is particularly useful for tracking solar granulation and magnetic fields. By computing cross-correlations in the Fourier space, it can be efficient and able to identify local displacements effectively \citep{Fisher:2008}.
\citet{Byrne:2013} discusses the FLCT sampling cadence problem in the context of numerical differentiation and highlights that na\"ive application of simple difference formulas fails in these cases. The core issue arises when data is sampled infrequently or irregularly, leading to errors in differentiation techniques. When sampling times are non-uniform or low-cadence, a more complex approach might be necessary, typically using higher-order polynomial fits or adaptive numerical schemes. This issue, which occurs with some types of heliospheric data, or when there are missing data frames, made us look for alternative ways of velocity field estimation \citep{Stepanyuk:2024}.  Optical flow is a fundamental concept in computer vision and image processing. It refers to the apparent motion of brightness patterns in the image; (the word `optical' comes with the name of this class of methods historically, there is no requirement for the data to belong to any particular part of the spectra) .


\begin{figure}[htp]
    \centering
    \includegraphics[width=16cm]{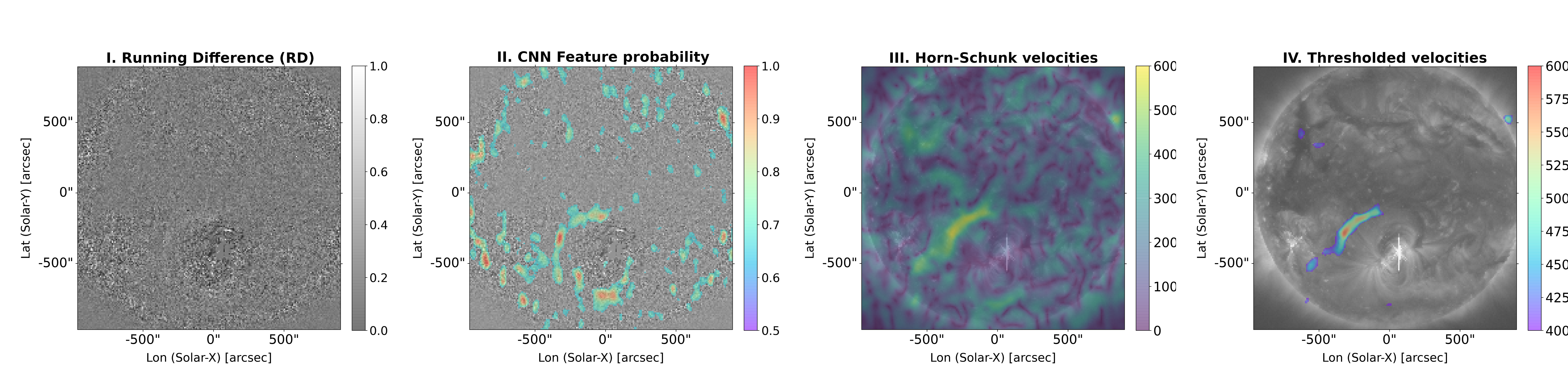}
    \caption {CBF velocity field calculation. The Horn–Schunk algorithm was applied to the results of the CNN image segmentation model. The initial Running Difference (RD) image and feature mask are presented in panels I and II. Panels III and IV correspond to the unfiltered and thresholded velocity fields. These images illustrate the idea that velocity field estimation techniques allow to perform image segmentation even with greater precision by distinguishing moving details.}
    \label{fig:velocity_panel}
\end{figure}

The Lucas-Kanade (LK) method estimates optical flow locally by assuming motion is constant within small windows, making it fast, robust to noise, and ideal for real-time feature tracking. However, it struggles with large displacements and homogeneous regions. The method works well for small motions and is more accurate in estimating finer details. \citep{lucas:1981}


The Horn-Schunck \citep{horn:1981}[HS] Method is global in nature, meaning it uses equations based on brightness constancy over the entire image. It introduces a smoothness constraint that penalizes non-smooth flow fields, aiming to capture more global motion. The regularization term (smoothness constraint) makes it robust to noise but can sometimes over-smooth the result, missing finer motion details. For the application to large-scale features such as CBFs, we find that using HS is preferable.





Figure \ref{fig:velocity_panel} shows the result of applying the HS algorithm to the May 10, 2022 CBF. The first panel shows an AIA running difference frame for context. In panel II, the CBF shape was captured well by the hybrid CNN model. Nevertheless, some pixel groups were also considered by the CNN as high probability pixels, while they clearly do not belong to the CBF. Such segmentation quality could be enough for human visual interpretation
of images, but may be problematic in automated pipelines. Panel III shows the velocity field obtained with Horn-Schunk algorithm. The EUV wave is more clearly distinguished from the static details with similar intensities and shapes. Finally, the thresholded velocity fields as calculated with the HS algorithm are presented in Panel IV in the figure. The average estimated velocity of $>$700 km/s for the CBF, obtained through this additional thresholding, is fully consistent with \citet{Rigney:2024}. Thus, the proposed method not only validates the velocity estimation by independent means, but also improves the detection and tracking of CBFs.

\subsection{Velocity estimation techniques: improving image segmentation performance}
Image-by-image segmentation, traditionally performed using algorithmic methods and more recently explored with data-driven approaches, often considers only the spatial aspects of image features, entirely overlooking their temporal evolution. Beyond their value in enhancing the qualitative understanding of physical processes, velocity field estimation techniques, such as FLCT (Fourier Local Correlation Tracking) and Optical Flow methods, can further help differentiate genuine details from spurious features as these methods operate within a space-time context rather than being limited to purely spatial processing. Simply put, capturing feature movement helps to distinguish moving details from static ones, even if they have similar intensities or shapes.
Figure \ref{fig:velocity_panel} illustrates this concept. The EUV wave shape was captured well by the model (panel II), nevertheless some pixel groups were also considered by the CNN as high probability pixels, while they clearly do not belong to the EUV wave. Such segmentation quality could be enough for human visual interpretation of images, but may be problematic in automatic pipelines. Panel III shows the velocity field obtained with Horn-Schunk algorithm. The EUV wave is more clearly distinguished from the static details with similar intensities and shapes.  

\section{Conclusions}
In this publication we discussed our general approach on finding optimal configuration and setup to perform data-driven segmentation of eruptive solar phenomena with CNNs. We have demonstrated its performance by training automatically generated networks of U-NET-like architecture and applying them to EUV solar observations. Training sets consisted of SDO/AIA 193~\AA-channel BD data, as well as synthetic-augmented images. Visual validation of the results shows that models developed within our approach perform reasonably well. According to the IoU metrics, the results are in good agreement with `ground-truth' data obtained by the Wavetrack code. Once trained models can be efficiently applied to a larger number events. 

At the \url{https://gitlab.com/iahelio/helios_cnn/} repository we provide a command-line Python utility for image segmentation using our CNNs designed to process images of solar eruptive phenomena. The repository also includes regularly updated and newly published models. For more detailed information and model descriptions, please follow the documentation. 

The proposed approach and models we develop is a step forward towards automatic data-driven segmentation in the heliophysics domain and may contribute to better understanding of the physics behind observed phenomena.

\section*{Acknowledgments}
This work was supported by the LOFAR-BG project of the National Roadmap for Research Infrastructure of Bulgaria, under contracts D01-362/14.12.2023 and D01-110/30.06.2025 with the Ministry of Education and Science, as well as by the project MOSAIICS, VIHREN National Research Program, under contract KP-06-DV-8/18.12.2019. SDO/AIA data is courtesy of NASA/SDO and the AIA, EVE, and HMI science teams. This research used the SunPy open source software package \citep{Sunpy:2020}.

\section*{Open Research}
SDO/AIA data are available through the SDO JSOC at \url{http://jsoc.stanford.edu/ajax/lookdata.html}.  Regularly updated and new models are available at \url{https://gitlab.com/iahelio/helios_cnn/}.  Datasets and models described in the results section of this paper can be also found at a Zendodo repository \citep{Stepanyuk_dataset:2024} 

\section*{Conflict of Interest Statement}
The authors have no conflicts of interest to disclose.

\end{document}